\documentclass[structabstract]{aa}
\usepackage{graphicx}
\usepackage{txfonts}
\usepackage{longtable}

\begin{document}
%
   \title{Planetary companions in K giants $\beta$ Cancri, $\mu$ Leonis, \\
   and $\beta$~Ursae~Minoris\thanks{Based on observations made with the BOES instrument on the 1.8m telescope at Bohyunsan Optical Astronomy Observatory in Korea.}}
   \author{B.-C. Lee,\inst{1}
          I. Han,\inst{1}
          M.-G. Park,\inst{2}
          D. E. Mkrtichian,\inst{3,4}
          A. P. Hatzes,\inst{5}
          \and
          K.-M. Kim \inst{1}
          }

   \institute{Korea Astronomy and Space Science Institute, 776,
		Daedeokdae-Ro, Youseong-Gu, Daejeon 305-348, Korea\\
	      \email{[bclee;iwhan;kmkim]@kasi.re.kr}
	    \and
	     Department of Astronomy and Atmospheric Sciences,
	     Kyungpook National University, Daegu 702-701, Korea\\
	      \email{mgp@knu.ac.kr}
        \and
        National Astronomical Research Institute of Thailand, Chiang Mai 50200, Thailand
        \and
        Crimean Astrophysical Observatory, Taras Shevchenko National University of Kyiv, Nauchny, Crimea, 98409, Ukraine\\
          \email{davidmkrt@gmail.com}
        \and
        Th{\"u}ringer Landessternwarte Tautenburg (TLS), Sternwarte 5, 07778 Tautenburg, Germany\\
	      \email{artie@tls-tautenburg.de}
             }

   \date{Received 5 September 2013 / Accepted 1 May 2014}


  \abstract
   {}
   {The aim of our paper is to investigate the low-amplitude and long-period variations in evolved stars with a precise radial velocity (RV) survey.
   }
   {The high-resolution, the fiber-fed Bohyunsan Observatory Echelle Spectrograph (BOES) was used from 2003 to 2013 for a radial velocity survey of giant stars as part of the exoplanet search program at Bohyunsan Optical Astronomy Observatory (BOAO).
   }
   {We report the detection of three new planetary companions orbiting the K giants $\beta$~Cnc, $\mu$~Leo, and $\beta$~UMi. The planetary nature of the radial velocity variations is supported by analyzes of ancillary data. The \emph{HIPPARCOS} photometry shows no variations with periods close to those in RV variations and there is no strong correlation between the bisector velocity span (BVS) and the radial velocities for each star. Furthermore, the stars show weak or no core reversal in Ca II H lines indicating that they are inactive stars. The companion to $\beta$ Cnc has a minimum mass of 7.8 $\it M_\mathrm{Jup}$ in a 605-day orbit with an eccentricity of 0.08. The giant $\mu$ Leo is orbited by a companion of minimum mass of 2.4 $\it M_\mathrm{Jup}$ having a period of 357 days and an eccentricity of 0.09. The giant $\beta$ UMi is a known barium star and is suspected of harboring a white dwarf or substellar mass companion. Its companion has a minimum mass of 6.1 $\it M_\mathrm{Jup}$, a period of 522 days, and an eccentricity $e$ = 0.19.
   }
   {}


   \authorrunning{B.-C. Lee et al.}
   \titlerunning{Planetary companions in K giants $\beta$ Cancri, $\mu$ Leonis, and $\beta$ Ursae Minoris}
   \maketitle
%

\section{Introduction}
To date there are  over 900 confirmed exoplanets that have been discovered by various methods. More than 60\% of these have been detected by the precise radial velocity (RV) technique. At first, the RV method concentrated on searching for exoplanets around  F, G, and K main-sequence stars because these have a plethora of narrow stellar lines that are amenable to RV measurements. Furthermore, dwarfs display much smaller intrinsic RV variations compared to evolved stars.

Evolved stars such as K giants also have suitable absorption lines for RV measurements and allow us to probe planets around stars with higher mass compared to the surveys around dwarf stars. Since the discovery of the first exoplanets around K giant stars (Hatzes \& Cochran 1993; Frink et al. 2002) about 50 exoplanets have been discovered orbiting giant stars. Not only do RV surveys of giant stars allow us to probe planet formation around more massive stars but they also aid in our understanding of how stellar evolution affects planetary systems.

In this paper, we present precise RV measurements of three evolved stars. In Sect. 2, we describe the observations. In Sect. 3, the stellar characteristics are determined. Radial velocity analysis and the nature of RV measurements for each star will be presented in Sect. 4. Finally, in Sect. 5, we discuss and summarize our results from this study.

%

\section{Observations}
All of the observations were carried out using the fiber-fed high-resolution Bohyunsan Observatory Echelle Spectrograph (BOES) at Bohyunsan Optical Astronomy Observatory in Korea. The BOES covers a wavelength region from 3500\,${\AA}$ to 10~500\,${\AA}$ with one exposure and the spectrum is distributed over $\sim$~85 orders. The attached camera is equipped with 2K $\times$ 4K CCD (pixel size of 15 $\mu$ $\times$ 15 $\mu$). For this survey we used a fiber with a core diameter  of 80 $\mu$. This projected to  1.1 arcsec on the sky and yielded a resolving power of  \emph{R} = 90 000.
To provide precise RV measurements an iodine absorption (I$_{2}$) cell was used. This provided absorption lines
suitable for wavelength calibration in the wavelength region of 4900$-$6000 {\AA}. The  estimated signal-to-noise (S/N) in the I$_{2}$ region was about 200 using a typical exposure time ranging from 6 minutes to 15 minutes.
The basic reduction of spectra was performed with the IRAF software package and DECH (Galazutdinov 1992) code.


\section{Stellar characteristics}
%

%
\begin{table*}
\begin{center}
\caption[]{Stellar parameters for the stars analyzed in the present paper.}
\label{tab1}
\begin{tabular}{lccccc}
\hline
\hline
    Parameter                &     & $\beta$ Cnc   & $\mu$ Leo       &  $\beta$ UMi  &  Reference     \\

\hline
    Spectral type      &     & K4 III  & K2 III   &  K4 IIIvar  & \emph{HIPPARCOS} (ESA 1997)  \\
    $\textit{$m_{v}$}$ &[mag]& 3.52    & 3.88     &   2.08      & \emph{HIPPARCOS} (ESA 1997)  \\
    $\textit{B-V}$     &[mag]& 1.481 $\pm$ 0.006 & 1.222 &   1.465 $\pm$ 0.005 & van Leeuwen (2007) \\
    age                &[Gyr]& 1.85 $\pm$ 0.34 & 3.35 $\pm$ 0.70 & 2.95 $\pm$ 1.03 & Derived\tablefootmark{a}  \\
    $\emph{d}$         &[pc]& 90.0 $\pm$ 7.7 & 40.9 $\pm$ 1.3  &   38.7  $\pm$ 0.8 & Famaey (2005) \\
                       &    & 92.99 $\pm$ 1.7 & 38.05 $\pm$ 0.23 & 40.14 $\pm$ 0.2 & Anderson \& Francis (2012)\\
    RV                 &[km s$^{-1}$]& 22.6 $\pm$ 0.2 & 13.6 $\pm$ 0.2 &   16.9 $\pm$ 0.3 & Gontcharov (2006) \\
    $\pi$              &[mas]& 10.75 $\pm$ 0.19 & 26.28 $\pm$ 0.16 &    24.91 $\pm$ 0.12 & van Leeuwen (2007) \\
    $T_{\rm{eff}}$     &[K]& 3990 $\pm$ 20 & 4436 $\pm$ 4  &  --  & Massarotti et al. (2008) \\
                       &   & 4150  & 4660  & 4150  & Lafrasse et al. (2010)  \\
                       &   & -- & 4453 $\pm$ 19 &   4067 $\pm$ 13  & Wu et al. (2011)  \\
                       &   & 4092.1 $\pm$ 17.5  & 4538.2 $\pm$ 27.5  & 4126.0 $\pm$ 25.0  & This work  \\
    $\rm{[Fe/H]}$      &[dex]& -- & 0.30 $\pm$ 0.04 &   -- 0.13 $\pm$ 0.05  & Wu et al. (2011)  \\
                       &     & -- 0.17 & 0.25 $\pm$ 0.02 &   -- 0.17  & Anderson \& Francis (2012)  \\
                       &     & -- 0.29 $\pm$ 0.06  & 0.36 $\pm$ 0.05  & -- 0.27 $\pm$ 0.07  & This work  \\
    log $\it g$        &[cgs]& 1.9  & 2.1 &  1.9  & Lafrasse et al. (2010)  \\
                       &     & --   & 2.61 $\pm$ 0.08 &   1.70 $\pm$ 0.11 & Wu et al. (2011)  \\
                       &     & 1.4 $\pm$ 0.1 & 2.4 $\pm$ 0.1 & 1.5 $\pm$ 0.1 & This work  \\
                       &     & 1.28 $\pm$ 0.04 & 2.47 $\pm$ 0.04 & 1.39 $\pm$ 0.06 & Derived\tablefootmark{a} \\
    $v_{\rm{micro}}$        &[km s$^{-1}$]& 1.8 $\pm$ 0.1  & 1.4 $\pm$ 0.1  & 1.7 $\pm$ 0.1 &  This work  \\
    $\textit{$R_{\star}$}$  &[$R_{\odot}$]& -- & 16.2 &   -- & Rutten (1987) \\
                            &             & 48.96 $\pm$ 4.23 & --& 42.06 $\pm$ 0.91 & Piau et al.  (2011) \\
                            &             & 47.2 $\pm$ 1.3  & 11.4 $\pm$ 0.2 & 38.3 $\pm$ 1.1 & Derived\tablefootmark{a} \\
    $\textit{$M_{\star}$}$  &[$M_{\odot}$]& 1.7 $\pm$ 0.1 & 1.5 $\pm$ 0.1 & 1.4 $\pm$ 0.2 & Derived\tablefootmark{a} \\
    $\textit{$L_{\star}$}$  &[$L_{\odot}$] & 785.7 & 62.62  &   537.07   & Anderson \& Francis (2012)   \\
    $v_{\rm{rot}}$ sin $i$  &[km s$^{-1}$] & 2.1   & 1.2  & 1.7  & de Medeiros \& Mayor (1999) \\
                            &              & 3.7 $\pm$ 0.8 & 2.4 $\pm$ 0.5 & 1.7 $\pm$ 1.4 & G{\l}{\c e}bocki \& Gnaci{\'n}ski (2005) \\
                            &              & 4.88 & 5.06 &   --  &  Hekker \& Mel{\'e}ndez (2007)  \\
                            &              & 6.9  & 4.5  &  --  & Massarotti et al. (2008) \\
   log $R^{'}_{\rm HK}$     &              &  --     & --       &  -- 4.68  & Cornide et al. (1992)  \\
    $P_{\rm{rot}}$ / sin $i$  &[days] & -- & 268     & --       & Rutten (1987) \\
                              &       & 346--1137 & 114--480 & 625--6457 & Derived\tablefootmark{b} \\
\hline

\end{tabular}
\end{center}
\tablefoottext{a}{Derived using the online tool (http://stev.oapd.inaf.it/cgi-bin/param}).
\tablefoottext{b}{See text}.
\end{table*}
The stellar characteristics are important in discerning the nature of any RV variations. In particular
evolved stars themselves can exhibit low-amplitude and long-period RV variations produced by rotation and surface activity such as spots, plages, or filaments.

The basic parameters were taken from the \emph{HIPPARCOS} satellite catalog (ESA 1997) and improved results for the parallaxes from van Leeuwen (2007). The giant star $\beta$~Cnc (HD~69267, HIP~40526, HR~3249) is a K4 III with a visual magnitude of 3.52 and a $\textit{B-V}$ color index of 1.481 mag. The giant $\mu$~Leo (HD~85503, HIP~48455, HR~3905) is a K2 III with a visual magnitude of 3.88 and a $\textit{B-V}$ color index of 1.222 mag. Lastly, the K4 II star $\beta$~UMi (HD~131873, HIP~72607, HR~5563) has a visual magnitude of 2.08 and a $\textit{B-V}$ color index of 1.465 mag. SIMBAD classifies it as a variable star.

The stellar atmospheric parameters were determined using the TGVIT (Takeda et al. 2005) code based on the Kurucz (1992) atmosphere models. We used 164 ($\beta$ Cnc), 155 ($\mu$ Leo), and 190 ($\beta$ UMi) equivalent width (EW) measurements of Fe~I and Fe II lines. Table 1 lists all our derived stellar parameters as well as those reported in the literature. Our values are consistent with the published values.

Stellar radii and masses were estimated from the stars'  positions in the color--magnitude diagram and using the theoretical stellar isochrones of Bressan et al. (2012). This is an updated version of the code used to compute stellar evolutionary tracks by the PADOVA group.
We also adopted a version of the Bayesian estimation method (J{\o}rgensen \& Lindegren 2005; da Silva et al. 2006) by using the determined values for $T_{\mathrm{eff}}$, $\mathrm{[Fe/H]}$, $m_{v}$, and $\pi$. Our estimated parameters yielded $R_{\star}$ = 47.2 $\pm$ 1.3 $R_{\odot}$, $M_{\star}$ = 1.7 $\pm$ 0.1 $M_{\odot}$ ($\beta$ Cnc); $R_{\star}$ = 11.4 $\pm$ 0.2 $R_{\odot}$, $M_{\star}$ = 1.5 $\pm$ 0.1 $M_{\odot}$ ($\mu$ Leo); and $R_{\star}$ = 38.3 $\pm$ 1.1 $R_{\odot}$, $M_{\star}$ = 1.4 $\pm$ 0.2 $M_{\odot}$ ($\beta$~UMi).

The stellar rotational period is important in identifying the nature of any long-period RV variations. Sometimes low-amplitude and long-period RV variations arise from rotational modulation of surface structure (Lee et al. 2008; 2012).
Results of rotational velocities depend strongly on the adopted templates and methods used for calibration of line broadening and thus it would be inappropriate to choose any one value as the `correct' one.
With these determinations we have a range of $v_{\mathrm{rot}}$ sin $i$ measurements of 2.1$-$6.9 ($\beta$~Cnc), 1.2$-$5.06 ($\mu$~Leo), and 0.3$-$3.1 km~s$^{-1}$ ($\beta$~UMi). Adopting a stellar radius of 47.2 $\pm$ 1.3 ($\beta$~Cnc), 11.4 $\pm$ 0.2 ($\mu$~Leo), and 38.3 $\pm$ 1.1 $R_{\odot}$ ($\beta$~UMi), we obtain the ranges for the upper limit of the rotation periods of 346$-$1137 ($\beta$~Cnc), 114$-$480 ($\mu$~Leo), and 625$-$6457 days ($\beta$~UMi).
The basic stellar parameters are summarized in Table~\ref{tab1}.

\section{Radial velocity analysis}

\subsection{Precise radial velocity analysis and period search}
Radial velocities were determined using  the RVI2CELL (Han et al. 2007) code which is based on the method by Butler et al. (1996) and Valenti et al. (1995). The stellar spectrum superposed with the I$_2$ spectrum is modeled as a product of a high-resolution I$_2$  and a stellar template spectrum convolved with an instrumental profile (IP). To obtain the stellar spectrum, we used a stellar template spectrum that was obtained by deconvolving a pure stellar spectrum with the spectrograph IP determined from a specrtrum
of a tungsten-halogen lamp (THL)  observed through the I$_2$ cell. The long-term stability of the BOES with the I$_{2}$ cell was demonstrated by observing the RV standard star $\tau$ Ceti, which was constant with an rms scatter (intrinsic uncertainty) of $\sim$ 7 m s$^{-1}$ (Lee et al. 2013). The resulting RV measurements are listed as online data (Tables~\ref{tab2}, ~\ref{tab3}, and ~\ref{tab4}).

In order to search for periodicity in observed RV data we performed the Lomb-Scargle periodogram analysis (Scargle 1982).
The significance threshold level is determined from the average power of the Lomb-Scargle periodogram for frequencies between 0 c~d$^{-1}$ and 0.03 c~d$^{-1}$ with a frequency step of 0.00001 c~d$^{-1}$. We also adopted the method of iterative sine wave fitting and subsequent removal of found frequencies (prewhitening) for searching variability.
\subsection{Orbital solutions}
\subsubsection{$\beta$ Cnc (HD 69267, HIP 40526, HR 3249)}
%

   \begin{figure}
   \centering
   \includegraphics[width=8cm]{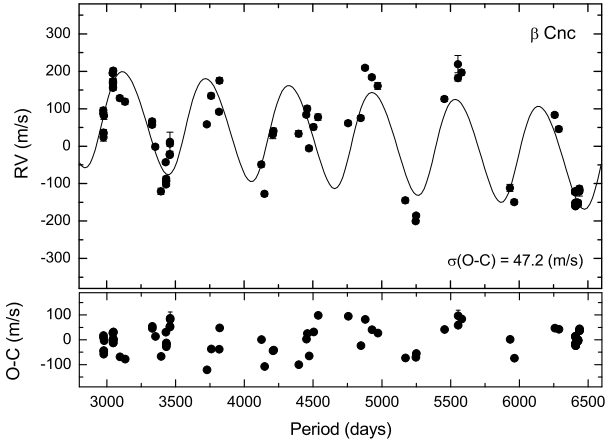}
      \caption{(\emph{top panel}) The RV measurements for $\beta$ Cnc from  December 2003 to May 2013. The solid line is the superposition of a linear trend with the orbital solution (see Table 2). (\emph{bottom panel}) The RV residuals after removing the orbital solution plus trend.
              }
         \label{rv1}
   \end{figure}
%
%
   \begin{figure}
   \centering
   \includegraphics[width=8cm]{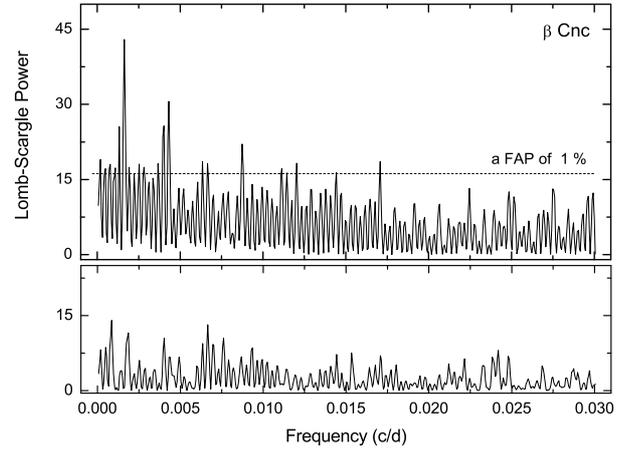}
      \caption{(\emph{top panel}) The Lomb-Scargle periodogram of the RV measurements for $\beta$~Cnc. The periodogram shows
      significant power at a frequency of 0.00165 c d$^{-1}$ corresponding to a period of 605.2 days. The horizontal line indicates a FAP threshold of 1\%. (\emph{bottom panel}) The periodogram of the RV residuals after removing the 605.2~d signal.
      }
         \label{power1}
   \end{figure}
%
%
   \begin{figure}
   \centering
   \includegraphics[width=8cm]{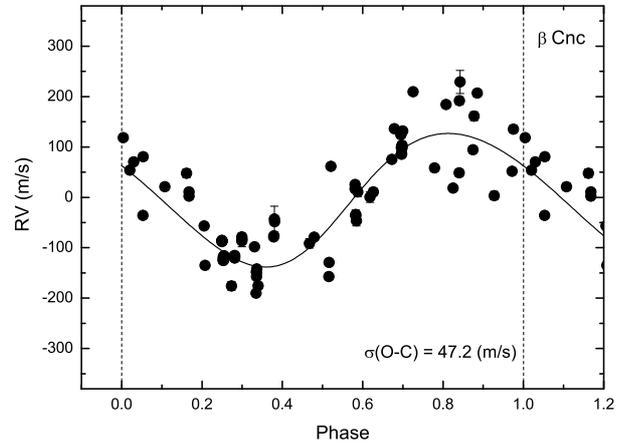}
      \caption{RV curve for $\beta$~Cnc phased to the orbital period of 605.2 days. The curve shows the orbital solution.
      }
         \label{phase1}
   \end{figure}
Hekker et al. (2008) conducted a precise RV survey of 179 K giants including $\beta$ Cnc using the CAT (\emph{R} = 60 000) at Lick observatory. They found a period of 673 days in the RV measurements of $\beta$ Cnc which is within $\sim$ 10\% of our measurements. They did not perform any analyzes of the spectral line profiles to determine the nature of the observed RV variations.

We monitored $\beta$ Cnc with the BOES for a time span of ten years, during which we acquired 85 RV measurements (Fig.~\ref{rv1}).
The RV variations show a slight linear trend of $-$3.1 $\times$ 10$^{- 2}$ m~s$^{-1}$~day$^{-1}$, possibly caused by a companion in a long-periodic orbit.
After removing the trend, the Lomb-Scargle periodogram of the RV measurements for $\beta$ Cnc shows a significant peak at a frequency of 0.00165 c d$^{-1}$ corresponding to a period of 605.2 days (Fig.~\ref{power1}).
The Lomb-Scargle power of this peak corresponds to a false alarm probability of (FAP) of 7 $\times$ 10$^{-7}$. This was verified using a bootstrap randomization process.
The RV data were shuffled 200\,000 times keeping the times fixed. In no instance did the Lomb-Scargle power of the random data exceed the power of the unshuffled data. This indicates that the FAP is $\ll$ 5$\times$10$^{-6}$.
An orbital fit to the RV data (curves in Figs. 3 and 5) yields  a period $P$ = 605.2 $\pm$ 4.0 days, a velocity amplitude $K$ = 133.0 $\pm$ 8.8 m~s$^{-1}$, and an eccentricity $e$ = 0.08 $\pm$ 0.02. Figure~\ref{phase1} shows the RV data phased to the orbital period. For the adopted stellar mass of 1.7 $\pm$ 0.1 for $\beta$ Cnc, it has a planetary companion of 7.8 $\pm$ 0.8 $\it M_\mathrm{Jup}$ at a separation of 1.7 $\pm$ 0.1 AU from the host star.
\subsubsection{$\mu$ Leo (HD 85503, HIP 48455, HR 3905)}
%

   \begin{figure}
   \centering
   \includegraphics[width=8cm]{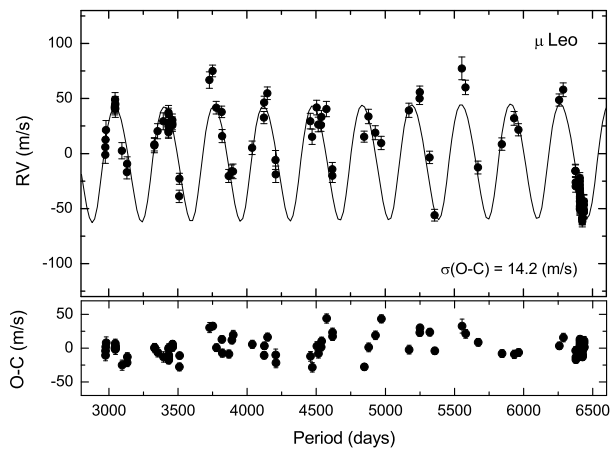}
      \caption{(\emph{top panel}) The RV measurements for $\mu$~Leo from December 2003 to May 2013. The solid line is the orbital solution with a period of 357.8 days and an eccentricity of 0.09. (\emph{bottom panel}) The RV residuals after removing the orbital solution.
              }
         \label{rv2}
   \end{figure}
%
%
   \begin{figure}
   \centering
   \includegraphics[width=8cm]{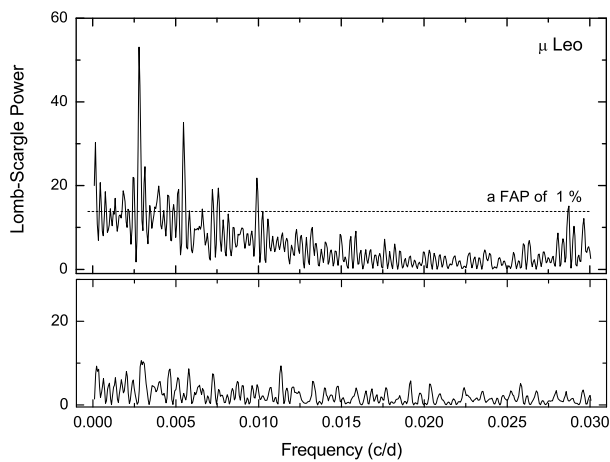}
      \caption{(\emph{top panel}) The Lomb-Scargle periodogram of the RV measurements for $\mu$~Leo. Significant power occurs at a frequency of 0.00279 c~d$^{-1}$ corresponding to a period of 357.8 days. The horizontal line indicates a FAP threshold of 1\%. (\emph{bottom panel}) The  periodogram of the RV residuals after subtracting the main period.
      }
         \label{power2}
   \end{figure}
%
%
   \begin{figure}
   \centering
   \includegraphics[width=8cm]{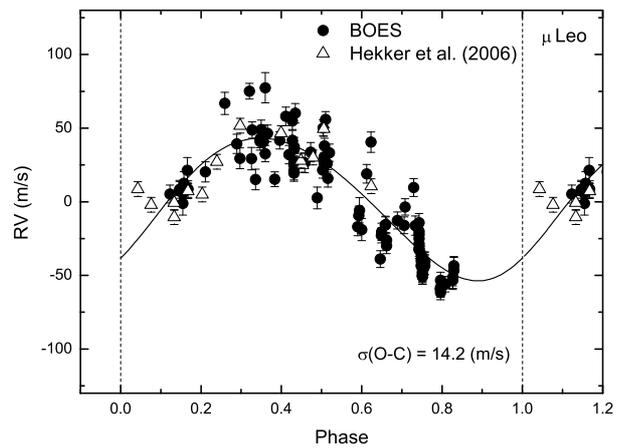}
      \caption{The RV measurements for $\mu$~Leo phased to the orbital period of 357.8 days. Open triangles indicate the measurements of the Hekker~et~al. (2006). The curve represents the orbital solution.
      }
         \label{phase2}
   \end{figure}
Between December 2003 and May 2013, we obtained 103 spectra for $\mu$ Leo (Fig.~\ref{rv2}).
In Fig.~\ref{power2}, the Lomb-Scargle periodogram of the data for $\mu$ Leo exhibits a dominant peak at a period of 357.8 $\pm$ 1.2 days ($f_{1}$ = 0.00279 c d$^{-1}$) and with a FAP $\sim$ 10$^{-8}$. A bootstrap with 200\,000 shuffles confirmed this low FAP.
The RV measurements for $\mu$~Leo including the orbital fit is plotted in Fig.~\ref{rv2} and the RV curve shows an obvious sinusoidal variation with a $K$ = 52.0 $\pm$ 5.4 m s$^{-1}$ and an $e$ = 0.09 $\pm$ 0.06 as shown in Fig.~\ref{phase2}.
Assuming a stellar mass of 1.5 $\pm$ 0.1 $M_{\odot}$ for $\mu$~Leo, we derived a minimum mass of a planetary companion $m$~sin~$i$ = 2.4 $\pm$ 0.4 $\it M_\mathrm{Jup}$ at a distance $a$ = 1.1 $\pm$ 0.1 AU from $\mu$~Leo.

Hekker et al. (2006) investigated RV variations of K giant stars with the high-resolution Echelle spectrograph CAT (\emph{R}~=~60~000) to find astrometric standard stars. They reported that $\mu$ Leo was constant to a level of $\sim$ 20 m s$^{-1}$ over a 1458-day interval and suggested using  it as a RV standard star. However, this result was based on only 14 measurements taken over four years. They proposed that stars with systematic RV variations or trends and with observed RV standard deviation of less than 20 m s$^{-1}$ be considered as constant stars. This adopted criterion is somewhat arbitrary.
We thus digitized the Hekker et al. (2006) data for $\mu$ Leo and made a simultaneous fit to the data allowing the zero point offset between the two to vary since they are independent data sets and neither is on an absolute RV scale. The resulting combined solution yields a period $P$ = 358.4 $\pm$ 1.1 days and a velocity amplitude $K$ = 44.5 $\pm$ 3.6 m~s$^{-1}$, values consistent with our initial solution.
We note that the period of the RV variations is more than six $\sigma$ away from one year. We have now included the Hekker et al. (2006) points phased to our orbital solution and they phase very well (Fig.~\ref{phase2}). The measurements of Hekker et al. (2006) only look inconsistent because of the poor phasing of the data and the near one-year orbital period.

\subsubsection{$\beta$ UMi (HD 131873, HIP 72607, HR 5563)}
%

   \begin{figure}
   \centering
   \includegraphics[width=8cm]{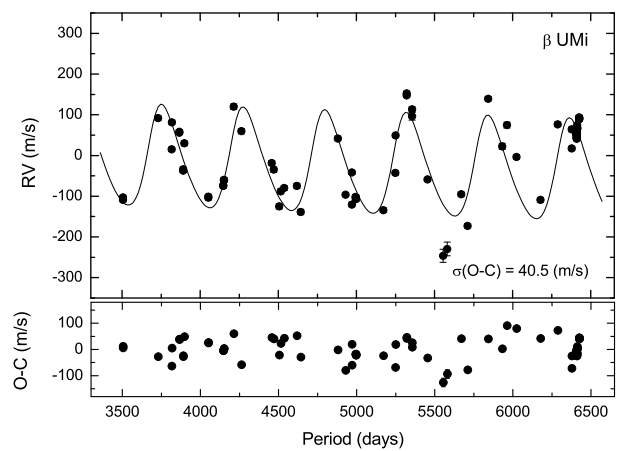}
      \caption{(\emph{top panel}) RV measurements for $\beta$ UMi from May 2005 to May 2013. The solid line is the superposition of a linear trend with the orbital solution with a period of 522.3 days and an eccentricity of 0.19. (\emph{bottom panel}) The
      RV residuals after subtracting the orbital solution plus trend.
              }
         \label{rv3}
   \end{figure}
%
%
   \begin{figure}
   \centering
   \includegraphics[width=8cm]{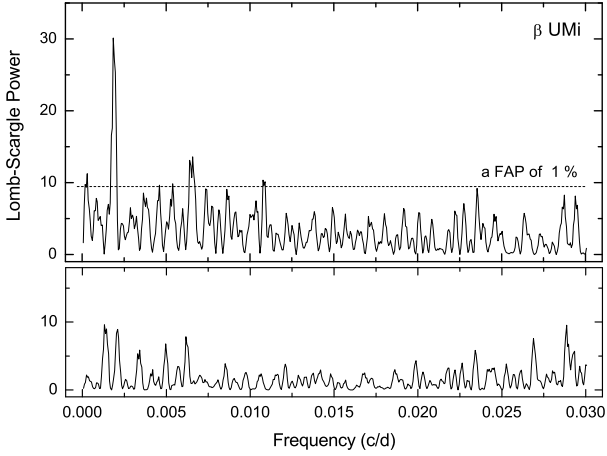}
      \caption{(\emph{top panel}) The Lomb-Scargle periodogram of the RV measurements for $\beta$ UMi. Significant power occurs
       at a frequency of 0.00191 c d$^{-1}$ corresponding to a period of 522.3 days. The horizontal line indicates a FAP threshold of 1\%. (\emph{bottom panel}) Periodogram of the RV residuals after subtracting the dominant period.
      }
         \label{power3}
   \end{figure}
%
%
   \begin{figure}
   \centering
   \includegraphics[width=8cm]{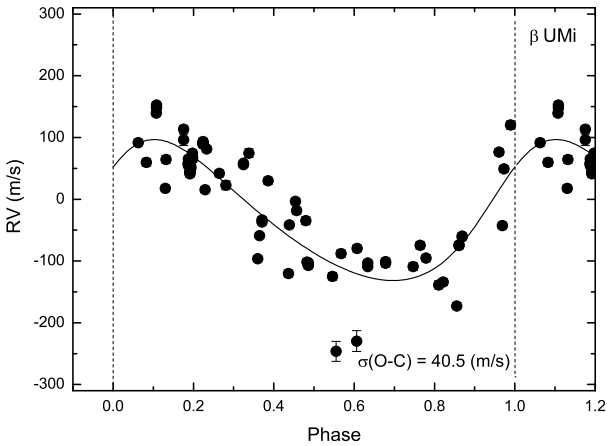}
      \caption{Phased RV curve for $\beta$ UMi with a period of 522.3 days. The curve represents the orbital solution.
      }
         \label{phase3}
   \end{figure}
%

%
\begin{table*}
\begin{center}
\caption{Orbital parameters for $\beta$ Cnc b, $\mu$ Leo b, and $\beta$ UMi b.}
\label{tab5}
\begin{tabular}{lcccc}
\hline
\hline
    Parameter                       &     & $\beta$ Cnc b        & $\mu$ Leo b          & $\beta$ UMi b       \\

\hline
    Period                       &[days] & 605.2  $\pm$ 4.0     & 357.8  $\pm$ 1.2     & 522.3 $\pm$ 2.7    \\
    $\it T$$_{\rm{periastron}}$  &[JD]   & 2453229.8 $\pm$ 9.8  & 2452921.0 $\pm$ 17.3  & 2453175.3 $\pm$ 4.6 \\
    $\it{K}$                     &[m s$^{-1}$] & 133.0  $\pm$ 8.8   & 52.0  $\pm$ 5.4  & 126.1  $\pm$ 8.1  \\
    $\it{e}$                     & & 0.08   $\pm$ 0.02    & 0.09  $\pm$ 0.06     & 0.19   $\pm$ 0.02 \\
    $\omega$                     &[deg] & 58.9   $\pm$ 5.9     & 227.1 $\pm$ 16.0     & 307.4   $\pm$ 2.8  \\
    slope           &[m s$^{-1}$ day$^{-1}$] & \textbf{--3.1 $\times$ 10$^{- 2}$}  & --  & \textbf{--1.3 $\times$ 10$^{- 2}$} \\
    $\sigma$ (O-C)               &[m s$^{-1}$]  & 47.2            & 14.2             & 40.5          \\
\hline
    with $\textit{$M_{\star}$}$  &[$M_{\odot}$]   & 1.7 $\pm$ 0.1   & 1.5 $\pm$ 0.1   & 1.4 $\pm$ 0.2  \\
    $m$ sin $i$                  &[$M_{\rm{Jup}}$]& 7.8 $\pm$ 0.8   & 2.4 $\pm$ 0.4   & 6.1 $\pm$ 1.0  \\
    $\it{a}$                     &[AU] & 1.7 $\pm$ 0.1    & 1.1 $\pm$ 0.1   & 1.4 $\pm$ 0.1   \\
\hline

\end{tabular}
\end{center}
\end{table*}
The giant $\beta$ UMi is classified as a barium star of type 0.3 (Lu 1991; Cornide et al. 1992).
Barium stars are spectral class G to K giants whose spectra indicate an overabundance of \emph{s}-process elements represented by the presence of singly ionized barium (Ba II) at 4554~{\AA} (Bidelman \& Keenan 1951).
Radial velocity studies suggested that all barium stars are in binary systems and white dwarf companions to some barium stars  have been detected with orbital periods ranging from 80 days to 10 years (McClure \& Woodsworth 1990). McClure et al. (1980) showed that $\beta$ UMi has a standard deviation of 0.54 km s$^{-1}$ based on nine spectra taken over 200 days.

We obtained  a total of 78 RV measurements for $\beta$ UMi between May 2005 and May 2013 (Fig.\ref{rv3}). The Lomb-Scargle periodogram of the data exhibits a dominant peak at a period of 522.3 days ($f_{1}$ = 0.00191 c d$^{-1}$) with a FAP $\approx$ 10$^{-5}$ (Fig.~\ref{power3}). A bootstrap using 200\,000 shuffles indicates that the FAP $<$ 5~$\times$~10$^{-6}$. An orbital solution yields $P$ = 522.3 $\pm$ 2.7 days, $K$~= 126.1 $\pm$ 8.1 m s$^{-1}$, and  $e$ = 0.19 $\pm$ 0.02. Figure~\ref{phase3} shows the RV data phase-folded to the orbital solution.
Assuming a stellar mass of 1.4 $\pm$ 0.2 $M_{\odot}$, a planetary companion is $m$ sin $i$ = 6.1 $\pm$ 1.0 $\it M_\mathrm{Jup}$ at a distance $a$ = 1.4 $\pm$ 0.1 AU from the host star. The orbital elements for the companions for all stars are listed in Table~\ref{tab5}.

We note that there is significant scatter about the orbital solution for each star (see Table~\ref{tab5}). These are most likely due to stellar oscillations. We have calculated the fundamental periods of radial pulsations, as well as the expected periods and amplitudes from solar-like oscillations using the relationships of Kjeldsen \& Bedding (1995) in Table~\ref{tab6}. The periods are far too short to be an explanation for the RV variations. However, the expected RV amplitudes from stellar oscillations are entirely consistent with the rms scatter of the RV about the orbits.
Thus, the large observed RV scatter about the orbital solution can easily be accounted for by these stellar oscillations. We note that $\mu$~Leo is expected to have the smallest RV amplitude attributable to stellar oscillations, and indeed has the smallest observed RV scatter.
%

%
\begin{table}
\begin{center}
\caption{Radial pulsation modes for $\beta$ Cnc, $\mu$ Leo, and $\beta$ UMi.}
\label{tab6}
\begin{tabular}{lcccc}
\hline
\hline
    Mode                              &               & $\beta$ Cnc  & $\mu$ Leo  & $\beta$ UMi   \\
\hline
    Fundamental period       & [days]        & 14    & 0.9   & 11  \\
    Pulsation period         & [days]        & 2.2   & 0.37  & 2.4 \\
    Pulsation RV amplitude   & [m s$^{-1}$]  & 108   & 10    & 90 \\
\hline
\end{tabular}
\end{center}
\end{table}
\subsection{Photometric variations}
In order to search for possible brightness variations that may be caused by the rotational modulation of large stellar spots, if any, the \emph{HIPPARCOS} photometry data obtained from March 1990 to October 1992 for $\beta$~Cnc, from December 1989 to December 1992 for $\mu$~Leo, and from December 1989 to March 1993 for $\beta$~UMi were analyzed. This database yields 82, 117, and 123 measurements for $\beta$~Cnc, $\mu$~Leo, and $\beta$~UMi, respectively. All stars were photometrically constant down to an rms scatter of 0.007 mag for  $\beta$~Cnc, 0.005 mag for $\mu$~Leo, and 0.008 mag for $\beta$~UMi. This  corresponds to variation over the time span of the observations of only 0.20\%, 0.13\%, and 0.38\% for the three stars, respectively. We should note that the \emph{HIPPARCOS} measurements were not contemporaneous with our RV measurements. Thus, it is possible that  stellar surface structures were present on the stars at the time of our observations, but were not present at the time of the \emph{HIPPARCOS} measurements.

Tarrant et al. (2008) analyzed a high-quality photometric time series of around 1000 days obtained from the Solar Mass Ejection Imager (SMEI) instrument on the Coriolis satellite. Their analysis showed a variability in $\beta$ UMi with a dominant period of $\approx$ 4.6 days. Two oscillation modes appear to be present at 2.44 and 2.92 $\mu$Hz, with a spacing of 0.48 $\mu$Hz. They  interpreted these as consecutive overtones of an acoustic spectrum.

\subsection{Line bisector analysis}
   \begin{figure}
   \centering
   \includegraphics[width=8cm]{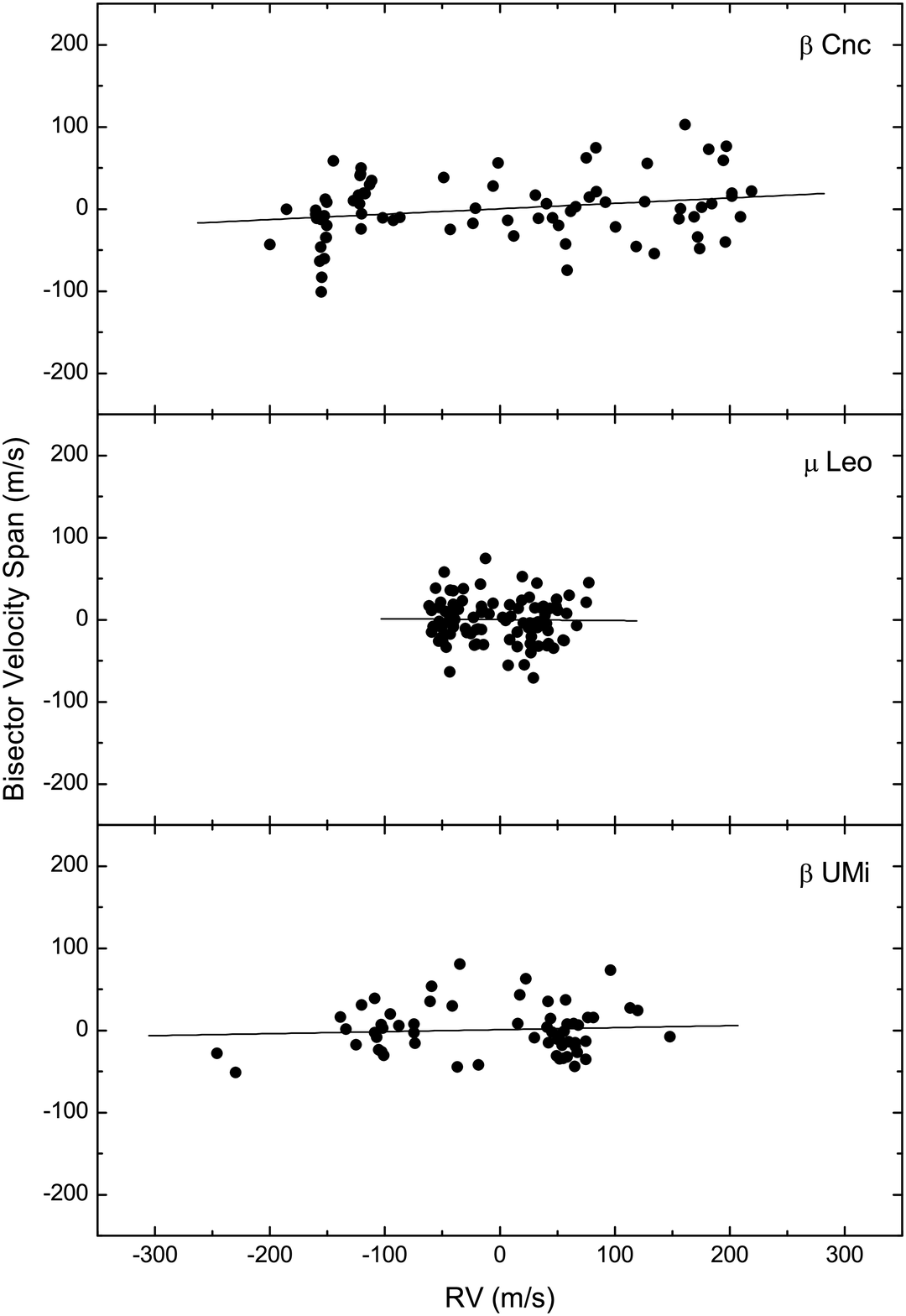}
      \caption{The BVS versus RV measurements  for $\beta$ Cnc, $\mu$ Leo, and $\beta$~UMi (\emph{top} to \emph{bottom panel}).
      The lines show the least-squares fit.
        }
        \label{BVS}
   \end{figure}
Stellar rotational modulations of inhomogeneous surface features can create variable asymmetries in the spectral line profiles. The differential RV measurements between the high and low flux levels of the line profile bisector are defined as a bisector velocity span (BVS). Line bisector variations showed that surface features were the cause of the RV variations in the main-sequence star HD 166435 (Queloz et al. 2001) and the K giant HD 78647 (Setiawan et al. 2004).

To search for variations in the spectral line shapes, the BVS was computed for a strong, unblended spectral features with a high flux level, namely Ni I 6643.6 {\AA} as described in Hatzes et al. (2005) and Lee et al. (2013). The selected line shows a high flux level and is located beyond the I$_{2}$ absorption region so contamination should not affect our bisector measurements.
We measured the BVS of the profile between two different flux levels with central depth levels of 0.75 and 0.4 as the span points, thereby avoiding the spectral core and wing where errors of the bisector measurements are large.
The BVS variations for $\beta$~Cnc, $\mu$~Leo, and $\beta$~UMi as a function of RV are shown in Fig.~\ref{BVS}. They do not show any obvious correlations, with the mean slopes of 0.06 ($\beta$~Cnc), $-$ 0.01 ($\mu$~Leo), and 0.04 ($\beta$~UMi). This suggests that RV variations are not caused by line-shape changes produced by rotational modulation of surface features.
\subsection{Chromospheric activities}
\begin{figure}
   \centering
   \includegraphics[width=8cm]{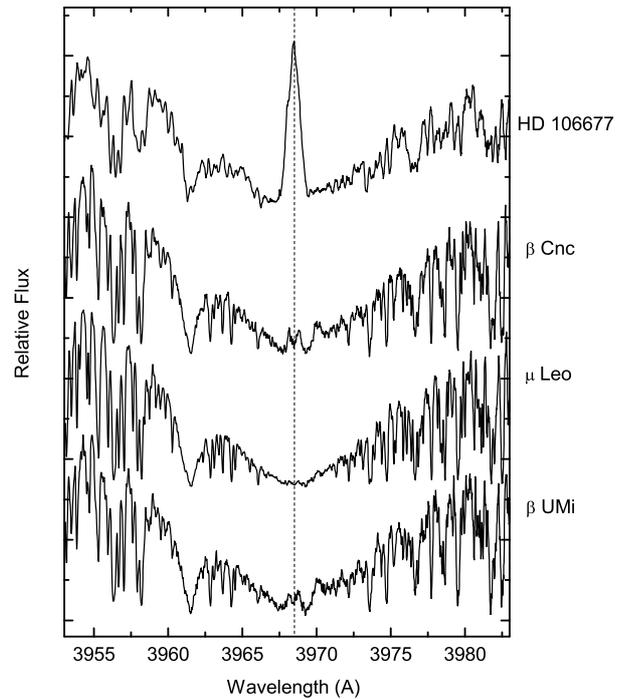}
      \caption{The Ca II H line for our program stars. $\mu$ Leo shows no significant core reversal in the line compared to that in the chromospheric active star HD 106677 (K0 III). There are weak central emissions in the center of the Ca II H line for $\beta$ Cnc and $\beta$ UMi. The vertical dotted line indicates the center of the Ca II H regions.
        }
        \label{Ca1}
   \end{figure}

Eberhart \& Schwarzschild (1913) were the first who discovered bright emission lines in the cores of the strong Ca II absorption features for $\alpha$~ Boo, $\alpha$~Tau, and $\sigma$~Gem. The emission at the line center implies that the source function in the chromosphere is greater than that in the photosphere. Stellar activity, such as spots, plages, and filaments, can induce RV variations that can mask or even mimic the RV signature of orbiting planetary companions.
Figure~\ref{Ca1} shows the Ca II H line region of the BOES spectra. The star $\mu$~Leo lacks a prominent core emission feature in Ca II H.
There is slight central emission in  $\beta$~Cnc and $\beta$~UMi indicating a low level of stellar activity.
Strassmeier et al. (1990) proposed that $\beta$ Cnc should not be considered chromospherically active based on  observations of the Ca II H \& K and H$\alpha$ region.

%
%


\section{Discussion and summary}
We find that $\beta$ Cnc, $\mu$ Leo, and $\beta$ UMi show evidence of low-amplitude and long-period RV variations.
The periodic RV variations in evolved stars may have been caused stellar pulsations, rotational modulations by inhomogeneous surface features, or planetary companions. Generally, the estimated period of the fundamental radial mode pulsations in early K giants is several days and pulsations may be well separated from other types of variability.

In order to distinguish between stellar activity and pure RV variations, we applied several diagnostic tools: Photometric data, Ca II H line inspection, rotational periods, and BVS measurements.
Above all we identified the \emph{HIPPARCOS} photometry and calculated periodograms for significant frequencies close to the obtained RV period. None of the stars shows photometric variations related to the observed RV variations.
Nevertheless, $\beta$~UMi is classified as a variable star in the SIMBAD database and it needs more investigations related to the RV variations.
Smith (2003) states that $\beta$ UMi exhibits little or no visual variation from the time-averaged DIRBE\footnote{The Diffuse Infrared Background Experiment (DIRBE) instrument on the Cosmic Background Explorer (COBE) satellite.} color-color diagram plots.
However, Tarrant et al. (2008) analyzed a high-quality photometric time series and showed a variability of short-period of $\approx$ 4.6 days.
Therefore, there are no long-period photometric variations comparable to RV period for $\beta$~UMi, even if it is a known variable star. We note that \emph{HIPPARCOS} data were not obtained contemporaneous with the spectral observations.

The examination of the activity indicator Ca II H line does not show any obvious evidence of chromospheric activity in $\mu$~Leo. While the weak central emissions are seen in the center of Ca~II~H line for $\beta$ Cnc and $\beta$ UMi, which are not strong enough to warrant being chromospherically active (Strassmeier et al. 1990; Cornide et al. 1992).

Giants typically have rotational periods of several hundred days and our samples show rotation periods (upper limit) of 346$-$1137 ($\beta$~Cnc), 114$-$480 ($\mu$~Leo), and 625$-$6457 days ($\beta$~UMi). Although there are no strong correlations between RV periods and rotational periods, we do not exclude the rotational modulations. We note that the detected RV periods could be compatible with the rotational periods of the samples, even if that is not the most likely explanation.

Our bisector analysis showed that the amplitude of variations  for the BVS are smaller than those of the RV.
The BOES exhibits a slightly unstable IP variation due to the possible mechanical, thermal and ambient air pressure instabilities from season to season. For RV measurements, most of the change in the IP is taken into account in I$_{2}$ cell calibration and the modelling process made for each spectral chunk, so this does not pose a serious problem and provides a long-term RV accuracy of $\sim$ 7 m~s$^{-1}$. The IP variations directly affect the line-bisector measurements, however, because of a differential character of the BVS measurements it has a reduced influence compared to the full amplitude of the IP variations. The short-term accuracy of the BVS measurements was also measured by comparison of several spectra obtained on the same night or during consecutive nights and is of $\sim$ 17 ($\beta$~Cnc), $\sim$ 8 ($\mu$~Leo), and $\sim$ 9 m s$^{-1}$ ($\beta$~UMi). The long-term accuracy of the BVS measurements might be larger than the short-term one.
However, as already mentioned, there are no correlations between the BVS and the RV variations slopes of 0.06 ($\beta$~Cnc), $-$ 0.01 ($\mu$~Leo), and 0.04 ($\beta$~UMi). Thus, this strongly suggests that RV variations arise from orbiting companion rather than rotational modulation of inhomogeneous surface features.

Thus, we conclude that K giants $\beta$ Cnc, $\mu$ Leo, and $\beta$ UMi host exoplanet companions with periods of 605, 357, and 522 days and that have minimum masses of 7.8, 2.4, and 6.1 $\it M_\mathrm{Jup}$, respectively.
%


\begin{acknowledgements}
      BCL acknowledges partial support by the KASI (Korea Astronomy and Space Science Institute) grant 2013-9-400-00. Support for MGP was provided by the National Research Foundation of Korea to the Center for Galaxy Evolution Research (No. 2012-0027910).
      DEM acknowledges his work as part of the research activity of the National Astronomical Research Institute of Thailand (NARIT), which is supported by the Ministry of Science and Technology of Thailand. APH acknowledges the support of DFG grant HA 3279/8-1. This research made use of the SIMBAD database, operated at the CDS, Strasbourg, France. 

\end{acknowledgements}
%


\Online

\begin{table}
\begin{center}
\caption{RV measurements for $\beta$ Cnc from December 2003 to May 2013 using the BOES.}
\label{tab2}
\begin{tabular}{cccccc}
\hline\hline

 JD         & $\Delta$RV  & $\pm \sigma$ &        JD & $\Delta$RV  & $\pm \sigma$  \\
 -2 450 000 & m\,s$^{-1}$ &  m\,s$^{-1}$ & -2 450 000  & m\,s$^{-1}$ &  m\,s$^{-1}$  \\
\hline

2976.280109 &      87.1  &      7.3 &   4505.159203  &     51.2 &       7.4  \\
2976.289798 &      95.5  &      7.4 &   4538.052625  &     77.7 &       8.5  \\
2977.335899 &      36.2  &      8.7 &   4755.373227  &     61.5 &       5.9  \\
2977.346016 &      33.8  &      7.1 &   4847.226746  &     75.3 &       6.4  \\
2978.225529 &      23.7  &     10.4 &   4879.106571  &    209.5 &       5.8  \\
2980.368679 &      80.9  &      9.4 &   4929.030885  &    184.3 &       5.8  \\
3045.112684 &     194.5  &      5.5 &   4970.990827  &    161.0 &       8.9  \\
3045.115774 &     196.1  &      5.7 &   5171.183837  &   -144.8 &       7.0  \\
3046.129161 &     168.9  &     10.5 &   5248.102651  &   -200.1 &       5.4  \\
3046.134219 &     155.8  &      6.7 &   5251.088009  &   -185.5 &       5.1  \\
3046.138930 &     157.2  &      6.9 &   5456.345312  &    126.2 &       6.5  \\
3047.133959 &     173.6  &      6.9 &   5554.324011  &    181.8 &       8.1  \\
3047.139260 &     168.1  &      6.6 &   5555.307049  &    219.1 &      22.9  \\
3047.145058 &     172.1  &      7.1 &   5581.136753  &    196.9 &       7.3  \\
3048.110284 &     201.7  &      6.3 &   5933.288218  &   -111.5 &       9.4  \\
3048.115122 &     201.7  &      6.2 &   5963.165961  &   -149.2 &       6.5  \\
3096.044410 &     128.3  &      6.8 &   6258.263659  &     83.6 &       6.0  \\
3133.008102 &     118.7  &      7.6 &   6288.084326  &     45.6 &       6.8  \\
3331.306788 &      57.4  &      6.9 &   6406.972849  &   -120.5 &       5.1  \\
3331.313930 &      66.1  &      6.6 &   6406.975534  &   -122.8 &       5.7  \\
3354.251889 &      -1.7  &      5.7 &   6406.978126  &   -120.7 &       5.5  \\
3395.288439 &    -120.7  &      8.5 &   6406.980719  &   -121.6 &       6.1  \\
3430.060599 &     -43.0  &      5.8 &   6406.983438  &   -122.1 &       5.7  \\
3433.086267 &     -93.2  &      5.2 &   6408.990558  &   -160.4 &       5.7  \\
3433.094970 &     -92.8  &      5.3 &   6408.993208  &   -159.4 &       5.3  \\
3433.159052 &    -102.0  &      5.3 &   6408.995789  &   -152.9 &       6.1  \\
3433.215125 &     -87.1  &      6.4 &   6408.998370  &   -160.1 &       5.6  \\
3459.030871 &     -23.5  &      5.8 &   6409.000951  &   -155.9 &       5.5  \\
3459.092671 &     -21.1  &      5.9 &   6409.964376  &   -152.0 &       6.0  \\
3459.978369 &      11.9  &     25.5 &   6409.966956  &   -150.7 &       5.2  \\
3460.105383 &       6.9  &      7.0 &   6409.969537  &   -152.9 &       5.5  \\
3729.118598 &      58.3  &      5.5 &   6409.972106  &   -155.9 &       6.2  \\
3759.175627 &     134.5  &      7.1 &   6409.974676  &   -156.5 &       6.4  \\
3818.022391 &      91.9  &      5.7 &   6425.976335  &   -151.1 &       6.1  \\
3819.986278 &     175.4  &      7.6 &   6425.978811  &   -151.6 &       6.1  \\
4125.164450 &     -49.0  &      6.9 &   6425.981045  &   -150.5 &       6.7  \\
4147.219409 &    -127.2  &      6.5 &   6425.983278  &   -155.0 &       6.1  \\
4209.031435 &      30.7  &     10.7 &   6425.985512  &   -155.5 &       5.8  \\
4214.016959 &      40.7  &      7.6 &   6436.969583  &   -121.4 &      11.8  \\
4396.360066 &      33.4  &      8.2 &   6436.972719  &   -118.4 &       6.8  \\
4452.371416 &      84.1  &      5.9 &   6436.975994  &   -117.3 &       6.7  \\
4458.330683 &     100.5  &      6.9 &   6436.979270  &   -113.3 &       7.2  \\
4472.134645 &      -5.8  &      6.3 &                &          &            \\

\hline

\end{tabular}
\end{center}
\end{table}
%

%
\begin{table}
\begin{center}
\caption{RV measurements for $\mu$ Leo from December 2003 to May 2013 using the BOES.}
\label{tab3}
\begin{tabular}{cccccc}
\hline\hline

 JD         & $\Delta$RV  & $\pm \sigma$ &        JD   & $\Delta$RV  & $\pm \sigma$  \\
 -2 450 000 & m\,s$^{-1}$ &  m\,s$^{-1}$ & -2 450 000  & m\,s$^{-1}$ &  m\,s$^{-1}$  \\
\hline

2976.319041  &     -1.1  &      7.9 &   4879.129708 &      33.7 &       6.5   \\
2976.327688  &      5.8  &      6.9 &   4929.079872 &      19.0 &       6.3   \\
2977.367437  &     12.5  &      7.9 &   4971.032098 &       9.6 &       6.1   \\
2980.386393  &     21.4  &      8.6 &   5171.244821 &      39.3 &       6.5   \\
3045.143480  &     41.0  &      5.9 &   5248.172182 &      50.1 &       5.5   \\
3045.148167  &     42.5  &      5.7 &   5250.209506 &      55.9 &       5.4   \\
3046.257371  &     49.1  &      6.3 &   5321.076505 &      -3.5 &       5.6   \\
3046.263922  &     43.7  &      6.8 &   5356.981352 &     -56.0 &       5.3   \\
3048.178401  &     40.9  &      8.0 &   5554.357408 &      77.4 &      10.3   \\
3048.185600  &     45.1  &      6.9 &   5581.158753 &      60.1 &       6.6   \\
3096.078205  &      2.6  &      7.4 &   5672.121022 &     -12.7 &       5.8   \\
3131.994682  &    -16.9  &      6.1 &   5843.329723 &       8.6 &       5.7   \\
3133.041480  &     -9.4  &      6.5 &   5933.305254 &      32.0 &       6.2   \\
3331.326982  &      7.3  &      6.5 &   5963.219135 &      21.6 &       5.5   \\
3331.334563  &      8.4  &      6.1 &   6258.387761 &      48.7 &       5.5   \\
3354.341567  &     20.3  &      6.9 &   6288.377849 &      58.0 &       6.3   \\
3395.310466  &     29.3  &      7.6 &   6376.996874 &     -15.6 &       5.3   \\
3433.046145  &     19.4  &      5.5 &   6377.000890 &     -15.8 &       6.0   \\
3433.055959  &     24.9  &      6.1 &   6378.022524 &     -26.1 &       4.9   \\
3433.149174  &     21.1  &      6.2 &   6378.028345 &     -29.9 &       5.3   \\
3433.207181  &     27.1  &      5.6 &   6406.998183 &     -22.3 &       4.9   \\
3433.252284  &     35.7  &      5.5 &   6407.002986 &     -31.7 &       5.5   \\
3433.288752  &     37.4  &      6.4 &   6407.006956 &     -32.7 &       5.7   \\
3460.089749  &     25.8  &      5.7 &   6407.010937 &     -28.5 &       5.3   \\
3460.145323  &     30.5  &      5.4 &   6407.014906 &     -25.1 &       5.5   \\
3460.186964  &     27.4  &      5.8 &   6409.007038 &     -39.1 &       5.9   \\
3510.030050  &    -38.9  &      5.7 &   6409.011007 &     -43.7 &       5.9   \\
3511.025083  &    -22.9  &      5.0 &   6409.014977 &     -43.2 &       5.6   \\
3729.302643  &     66.8  &      7.5 &   6409.018946 &     -36.2 &       6.1   \\
3751.399694  &     75.1  &      5.3 &   6409.022916 &     -40.6 &       6.1   \\
3778.275652  &     41.7  &      5.8 &   6409.981114 &     -47.6 &       5.6   \\
3818.088849  &     37.9  &      5.3 &   6409.985083 &     -49.6 &       5.3   \\
3821.074419  &     15.9  &      6.0 &   6409.989041 &     -48.5 &       5.4   \\
3869.022756  &    -20.4  &      5.9 &   6409.992999 &     -50.8 &       5.1   \\
3889.028748  &    -16.0  &      5.7 &   6409.996969 &     -50.6 &       5.4   \\
3899.007779  &    -16.2  &      6.6 &   6412.105429 &     -40.7 &       5.0   \\
4038.358344  &      5.2  &      6.1 &   6412.112870 &     -40.4 &       6.4   \\
4123.155987  &     32.7  &      5.5 &   6412.120312 &     -40.5 &       6.2   \\
4125.181340  &     46.5  &      5.6 &   6412.127765 &     -43.6 &       6.0   \\
4147.239492  &     54.8  &      5.8 &   6412.135206 &     -43.0 &       6.4   \\
4207.116756  &     -5.8  &      8.6 &   6425.991361 &     -53.3 &       5.5   \\
4209.076642  &    -18.9  &      7.5 &   6425.993594 &     -58.2 &       5.7   \\
4458.350129  &     29.5  &      7.1 &   6425.995828 &     -59.2 &       5.6   \\
4472.241093  &     15.2  &      7.0 &   6425.998061 &     -59.4 &       5.2   \\
4505.204676  &     41.8  &      6.6 &   6426.000295 &     -61.5 &       5.1   \\
4516.263224  &     26.4  &      6.7 &   6437.008771 &     -52.8 &       6.0   \\
4536.128341  &     25.9  &      5.7 &   6437.012151 &     -51.3 &       5.8   \\
4538.123663  &     33.3  &      5.9 &   6437.015426 &     -53.3 &       6.2   \\
4575.075315  &     40.5  &      6.9 &   6437.991873 &     -43.4 &       5.6   \\
4618.010069  &    -14.2  &      6.2 &   6437.995947 &     -43.4 &       6.2   \\
4618.015428  &    -20.2  &      6.1 &   6437.999917 &     -46.7 &       5.5   \\
4847.264547  &     15.3  &      5.0 &               &           &             \\

\hline

\end{tabular}
\end{center}
\end{table}
%

%
\begin{table}
\begin{center}
\caption{RV measurements for $\beta$ UMi from May 2005 to May 2013 using the BOES.}
\label{tab4}
\begin{tabular}{cccccc}
\hline\hline

 JD         & $\Delta$RV  & $\pm \sigma$ &        JD   & $\Delta$RV  & $\pm \sigma$  \\
 -2 450 000 & m\,s$^{-1}$ &  m\,s$^{-1}$ & -2 450 000  & m\,s$^{-1}$ &  m\,s$^{-1}$  \\
\hline

3506.172953  &   -103.1  &      4.7 &   5356.176043  &    113.5 &       6.0  \\
3506.177015  &   -109.0  &      4.2 &   5454.959914  &    -58.8 &       4.8  \\
3730.295554  &     91.7  &      5.7 &   5554.391505  &   -246.2 &       16.1 \\
3817.086568  &     15.4  &      5.2 &   5581.173457  &   -229.7 &       16.8 \\
3819.228095  &     81.4  &      5.5 &   5671.221520  &    -95.3 &       5.6  \\
3867.076518  &     56.1  &      4.6 &   5711.201817  &   -172.7 &       5.2  \\
3867.081946  &     58.1  &      5.0 &   5842.934355  &    139.5 &       5.2  \\
3891.136052  &    -33.7  &      5.5 &   5933.327285  &     22.6 &       7.2  \\
3891.146005  &    -37.0  &      4.7 &   5963.349721  &     74.7 &       7.0  \\
3899.140711  &     29.9  &      5.7 &   6023.975198  &     -3.5 &       5.5  \\
4051.907979  &   -103.4  &      6.3 &   6176.934832  &   -108.9 &       6.2  \\
4051.912284  &   -101.0  &      5.9 &   6288.393855  &     76.5 &       5.7  \\
4147.285435  &    -73.8  &      5.5 &   6377.050930  &     17.5 &       5.4  \\
4147.286812  &    -74.8  &      5.4 &   6378.031452  &     64.4 &       6.0  \\
4151.353726  &    -60.8  &      5.2 &   6407.064962  &     55.1 &       5.4  \\
4151.356284  &    -59.6  &      5.7 &   6407.066663  &     58.4 &       5.6  \\
4214.100517  &    120.1  &      6.7 &   6407.068202  &     57.3 &       5.7  \\
4263.044404  &     59.8  &      6.3 &   6407.069742  &     65.2 &       5.4  \\
4458.373594  &    -18.5  &      5.9 &   6407.071293  &     58.5 &       5.6  \\
4470.399194  &    -34.7  &      6.4 &   6409.079881  &     44.0 &       5.3  \\
4505.254500  &   -125.1  &      6.2 &   6409.081374  &     51.4 &       5.7  \\
4516.335491  &    -88.0  &      6.1 &   6409.082566  &     42.5 &       5.4  \\
4537.248642  &    -79.8  &      5.8 &   6409.083758  &     41.2 &       5.5  \\
4619.026542  &    -74.6  &      6.3 &   6409.084950  &     45.3 &       5.0  \\
4643.123869  &   -138.8  &      6.6 &   6410.040317  &     50.3 &       5.7  \\
4880.229995  &     41.8  &      5.7 &   6410.041509  &     52.0 &       5.2  \\
4930.162805  &    -96.5  &      4.8 &   6410.042689  &     49.4 &       5.2  \\
4970.096160  &   -120.3  &      5.8 &   6410.043870  &     51.3 &       5.4  \\
4971.211001  &    -41.6  &      5.1 &   6410.045050  &     54.2 &       5.2  \\
4994.149858  &   -101.7  &      5.4 &   6412.141610  &     65.4 &       5.3  \\
4996.061933  &   -107.2  &      5.3 &   6412.143497  &     67.5 &       5.2  \\
4996.063669  &   -102.9  &      5.5 &   6412.145383  &     68.1 &       5.3  \\
4996.065115  &   -105.4  &      5.5 &   6412.147270  &     65.7 &       5.5  \\
5171.393525  &   -134.1  &      5.7 &   6412.149156  &     74.7 &       5.4  \\
5248.216130  &    -42.5  &      5.7 &   6426.006397  &     93.4 &       5.5  \\
5250.218252  &     49.2  &      5.4 &   6426.007936  &     90.5 &       5.1  \\
5321.109424  &    152.7  &      5.4 &   6426.009302  &     92.2 &       5.1  \\
5321.111125  &    147.8  &      5.3 &   6426.010494  &     88.9 &       4.9  \\
5356.123325  &     96.2  &      9.3 &   6426.011686  &     89.6 &       4.5  \\

\hline

\end{tabular}
\end{center}
\end{table}
%


\end{document}